\documentclass[
preprint,
 ,secnumarabic%
,amssymb, amsmath,nobibnotes, aps, prl,showpacs]{revtex4}
\usepackage{docs}%
\usepackage{bm}%
\usepackage{amsfonts,amsmath,amsopn}
\usepackage{isolatin1}
\usepackage{graphicx}
\usepackage[german]{varioref}
\usepackage[dvips]{epsfig}
\usepackage{graphics}
\usepackage[dvips]{color}
\usepackage{color}
\usepackage{colordvi}
\usepackage{dsfont}
\usepackage{psfrag}
\expandafter\ifx\csname package@font\endcsname\relax\else
 \expandafter\expandafter
 \expandafter\usepackage
 \expandafter\expandafter
 \expandafter{\csname package@font\endcsname}%
\fi
\begin{document}
\newcommand{\captionfonts}{\small}
\newcommand{\be}{\begin{eqnarray}}
\newcommand{\ee}{\end{eqnarray}}
\def\p#1#2{|#1\rangle \langle #2|}
\def\ket#1{|#1\rangle}
\def\bra#1{\langle #1|}
\def\refeq#1{(\ref{#1})}
\def\tb#1{{\overline{{\underline{ #1}}}}}
\def\im{\mbox{Im}}
\def\re{\mbox{Re}}
\def\nn{\nonumber}
\def\t{\mbox{tr}}
\def\sgn{\mbox{sgn}}
\def\Li{\mbox{Li}}
\def\P{\mbox{P}}
\def\d{\mbox d}
\def\i{\int_{-\infty}^{\infty}}
\def\ip{\int_{0}^{\infty}}
\def\mi{\int_{-\infty}^{0}}
\def\A{\mathfrak A}
\def\AA{{\overline{{\mathfrak{A}}}}}
\def\a{\mathfrak a}
\def\aa{{\overline{{\mathfrak{a}}}}}
\def\B{\mathfrak B}
\def\BB{{\overline{{\mathfrak{B}}}}}
\def\b{\mathfrak b}
\def\bb{{\overline{{\mathfrak{b}}}}}
\def\R{\mathcal R}
\def\dm{\mathfrak d}
\def\dd{{\overline{{\mathfrak{d}}}}}
\def\D{\mathfrak D}
\def\DD{{\overline{{\mathfrak{D}}}}}
\def\c{\mathfrak c}
\def\cc{{\overline{{\mathfrak{c}}}}}
\def\C{\mathcal C}
\def\CC{{\overline{{\mathfrak{C}}}}}
\def\Or{\mathcal O}
\def\F{\mathcal F_k}
\def\N{\mathcal N}
\def\I{\mathcal I}
\def\S{\mathcal S}
\def\G{\Gamma}
\def\L{\Lambda}
\def\la{\lambda}
\def\g{\gamma}
\def\al{\alpha}
\def\s{\sigma}
\def\e{\epsilon}
\def\k{\kappa}
\def\om{\omega}
\def\ve{\varepsilon}
\def\te{\text{e}}
\def\rmi{\text{i}}
\def\max{\text{max}}
\def\str{\text{str}}
\def\tr{\text{tr}}
\def\tC{\text C}
\def\dn{\text{dn}}
\def\Fo{\mathcal{F}_{1,k}}
\def\Ft{\mathcal{F}_{2,k}}
\def\vs{\varsigma}
\def\l{\left}
\def\r{\right}
\def\up{\uparrow}
\def\down{\downarrow}
\def\u{\underline}
\def\ov{\overline}
\title{Exact thermodynamic limit of short-range correlation functions of the antiferromagnetic $XXZ$-chain at finite temperatures}

\author{Michael Bortz\footnote{Address from April 10, 2005: Department of Theoretical Physics RSPhysSE, The Australian National University, Canberra ACT 0200, email: mib105@rsphysse.anu.edu.au}, Frank G\"ohmann}
%\email{bortz@fkt.physik.uni-dortmund.de}
\affiliation{Bergische Universit\"at Wuppertal, Fachbereich C $-$ Physik, 42097 Wuppertal, Germany}
\date{\today}%
%\tableofcontents

%\section{Introduction}
\begin{abstract}
We evaluate numerically certain multiple integrals representing nearest and next-nearest neighbor correlation functions of the spin-1/2 $XXZ$ Heisenberg infinite chain at finite temperatures.
\end{abstract}
\maketitle

\section{Introduction}
\label{intsec}
The determination of correlation functions is an important task both from a theoretical and from an experimental point of view. In scattering experiments correlation functions are measured, and detailed insight into the microscopic structure of the sample can be obtained. Scattering experiments may thus well decide about the adequateness of a given (simplified) theoretical model, once its correlations functions are calculated. On the other hand, taken for granted that a theoretical model adequately describes a certain sample, the knowledge of its correlation functions is useful to estimate the quality of the experimental setup. The principal problems on the theoretical side are threefold: To deal with many-body interactions between the particles, to account for finite temperatures and to carry out the thermodynamic limit of a macroscopic sample size. With regard to these intricate difficulties, techniques which allow for the exact calculation of correlation functions of at least some typical interacting quantum systems in the thermodynamic limit are highly welcome. 

Progress into this direction has recently been achieved by one of the authors in collaboration with A. Kl\"umper and A. Seel (GKS) \cite{gks1,gks2,gks3}. In these works, the one-dimensional  spin-1/2 $XXZ$ Heisenberg chain with periodic boundary conditions in a magnetic field,
\be
H=J \sum_{j=1}^L\l[S_j^xS_{j+1}^x + S_j^yS_{j+1}^y+\Delta \l(S^z_jS^z_{j+1}-\frac14\r)\r]-h\sum_{j=1}^L S^z_j\label{defh}\;,
\ee
was considered. Formulae for the correlation functions were derived by combining the quantum-transfer-matrix (QTM) approach \cite{suz85,suz87,kl92rsos,kl93} to the thermodynamics of integrable systems with certain results for matrix elements of the $XXZ$-chain \cite{kor82,sla89} which proved to be useful before \cite{kit00,kit02} at $T=0$. In the QTM approach, the free energy of the system is expressed in terms of an auxiliary function defined in the complex plane \cite{kl93}. This auxiliary function is determined uniquely as a solution of a non-linear integral equation. GKS found that essentially the same auxiliary function can be used to calculate $m$-point correlation functions for arbitrary temperatures $T$, again in the thermodynamic limit. As a result, all the correlation functions of spatial range $m$ are given in terms of $m$-fold integrals over combinations of the auxiliary function taken at different integration variables. 

In this work, the integral equations are solved numerically for all temperatures at $h=0$ and the multiple integrals are calculated for $m=2,3$, where we restrict ourselves to $\Delta\geq 1$ for $m=3$. In particular, we focus on $\langle S_j^x S_{j+m-1}^x\rangle_T$ and on the emptiness formation probability $P(m)$. In order to estimate the quality of our data we compare them with other exact results: first of all the nearest-neighbor correlators can be obtained independently of the multiple integral representation by taking the derivative of the free energy with respect to $\Delta$. Furthermore, very recently \cite{tsu05}  a high temperature expansion of the emptiness formation probability $P(3)$ at $\Delta=1$ has been performed up to the order $(J/T)^{42}$, starting from the multiple integral representation. At $T=0$, closed expressions are available \cite{tak77,boo01,boo03,boo05,kat03,kat04} as well. There are, moreover, the $XX$-limit ($\Delta\to0$) and the Ising limit ($J\to 0,\;\Delta\to\infty,\; J\Delta$ fixed) where the correlation functions are known over the whole range of temperatures \cite{mcc68,shi01,bax82}. All these independent results provide useful tests for the accuracy of our numerics. The main error we observe is due to the discretization of the integrals, an error which, in principle, can be made arbitrarily small by increasing the numerical accuracy.  

Besides the expected crossover behaviour from low to high temperatures, we find a surprising feature in $\langle S_j^x S_{j+m-1}^x\rangle_T$ for $\Delta>1$: the antiferromagnetic correlation is maximal at an intermediate temperature $T_0$, and not, as naively expected, at $T=0$. We explain this behaviour from a competition between quantum and thermodynamical fluctuations.

This paper is organized as follows: in the next section we give some details about the numerical procedure. Results are presented separately in the third section. The article ends with a summary and an outlook.
%%%%%%%%%%%%%%%%%%%%%%%%%%%%%%%%%%%%
%%Numerical procedure
%%%%%%%%%%%%%%%%%%%%%%%%%%%%%%%%%%%
\section{Numerical procedure}
\label{num}
Let $A_{1 \ldots m}$ be an operator which acts on sites $1,\ldots,m$ of the spin chain. In order to calculate $\langle A_{1 \ldots m}\rangle_T$, the operator $A_{1 \ldots m}$ is expanded in terms of the elementary matrices $e_\al^\beta$, 
\be
e_1^1=\l(\begin{array}{cc}1 & 0\\0&0\end{array}\r)\,,\;
e_1^2=\l(\begin{array}{cc}0 & 1\\0&0\end{array}\r)\,,\;
e_2^1=\l(\begin{array}{cc}0 & 0\\1&0\end{array}\r)\,,\;
e_2^2=\l(\begin{array}{cc}0 & 0\\0&1\end{array}\r)\,,\nn
\ee
such that 
\be
A_{1 \ldots m}=A_{\al_1\ldots\al_m}^{\beta_1\ldots\beta_m} e^{\;\;\;\al_1}_{1\,\beta_1}\,e^{\;\;\;\al_2}_{2\,\beta_2}\,\ldots\,\,e^{\;\;\;\al_m}_{m\,\beta_m}\nn.
\ee
Then it suffices to calculate the expectation value $\langle e_{1\;\beta_1}^{\;\;\;\al_1}\ldots e_{m\;\beta_m}^{\;\;\;\;\al_m}\rangle_T$ which defines the density matrix of the chain segment consisting of sites $1,\ldots,m$. 

In \cite{gks3} the following multiple integral representation for the density matrix was has been obtained \footnote{Eq. \refeq{gen} was formulated as a conjecture in \cite{gks3} and a proof was only sketched. By now a complete proof is available for the general case \cite{has05}.}:
\be
\langle e_{1\;\beta_1}^{\;\;\;\al_1}\ldots e_{m\;\beta_m}^{\;\;\;\;\al_m}\rangle_T&=&
\l[\prod_{j=1}^{|\al^+|}\int_{\C}\frac{\d \om_j}{2\pi\rmi (1+\a(\om_j))}\r]\l[\prod_{j=|\al^+|+1}^{m}\int_{\C}\frac{\d \om_j}{2\pi\rmi (1+\aa(\om_j))}\r]\nn\\
& &\times\frac{(-1)^m\det_m\l[\frac{1}{(b-1)!}\frac{\partial^{b-1}}{\partial \xi^{b-1}}G\l(\om_a,\xi\r)|_{\xi=0}\r]_{ab}}{\prod_{1\leq j<k\leq m} \sinh\l(\om_k-\om_j+\eta\r)}\nn\\
& &\times\l[\prod_{j=1}^{|\al^+|}\sinh^{\al_j^+-1}(\om_{|\al^+|-j+1}-\eta)\sinh^{m-\al_j^+}\om_{|\al^+|-j+1}\r]\nn\\
& &\times\l[\prod_{j=1}^{m-|\al^+|}\sinh^{\beta_j^--1}(\om_{|\al^+|+j}+\eta)\sinh^{m-\beta_j^-}\om_{|\al^+|+j}\r]\label{gen}\; .
\ee
Here $(\al_n)^m_{n=1}$ and $(\beta_n)^m_{n=1}$ denote sequences of up- and down-spins, where $1\equiv \up$ and $2\equiv \down$. The position of the $j$th up-spin (down-spin) in the sequence $(\al_n)^m_{n=1}$ ($(\beta_n)^m_{n=1}$) is $\al_j^+$ ($\beta_j^-$) and the number of up- (down-) spins in the corresponding sequences is defined as $|\al^+|$ ($|\beta^-|$). The contour $\C$ in \refeq{gen} is given in \cite{gks1}; it depends on the value of $\Delta=:\cosh \eta$. The parameter $\eta$ may be purely real and positive (massive case, $\Delta>1$), or imaginary with $\eta=:\rmi \g$, $\pi/2\geq\g\geq 0$ (massless case, $0\leq \Delta\leq 1$). The auxiliary functions $\a, \bar \a, G$ which occur in the integrand in \refeq{gen} are solutions of the following integral equations:
\be
\ln \a(\la)&=&-\beta h- \frac{J\beta \sinh\eta}{2}\, \hat\kappa_{\frac{\eta}{2}}(\la+\eta/2)-\int_\C\frac{\d \om}{2\pi\rmi}\hat\kappa_\eta(\la-\om)\ln(1+\a(\om))\label{lna}\\
G(\la,\xi)&=&  \hat\kappa_{\frac{\eta}{2}}(\la-\xi-\eta/2)+\int_\C\frac{\d \om}{2\pi\rmi} \,  \hat\kappa_\eta(\la-\om)\frac{G(\om,\xi)}{1+\a(\om)} \label{g}\\
\hat \kappa_\eta(\la)&=& \frac{\sinh( 2\eta)}{ \sinh(\la+\eta)\sinh(\la-\eta)}\nn\\
\aa&:=&1/\a\nn\,,
\ee
with $\beta:=1/T$. Note that the $T$- and $h$-dependence of the correlation functions enters through the auxiliary functions, whereas the $m$-dependence is encoded in the $m$-fold integral. Thus the evaluation of \refeq{gen} requires two steps: solving equations \refeq{lna}, \refeq{g} and then calculating the multiple integrals. In this work, the following correlation functions will be calculated numerically at $h=0$:
\be
 P(2) &:=&\langle e_{j\;1}^{\;\;\;1}e_{j+1\;1}^{\;\;\;\;\;\;\;1}\rangle_T=\frac14+\langle S^z_jS^z_{j+1}\rangle_T\nn\\
\langle S^x_jS^x_{j+1}\rangle_T &=&\frac12\langle e_{j\;1}^{\;\;\;2}e_{j+1\;2}^{\;\;\;\;\;\;\;1}\rangle_T\nn\\
 P(3) &:=&\langle e_{j\;1}^{\;\;\;1}e_{j+1\;1}^{\;\;\;\;\;\;\;1}e_{j+2\;1}^{\;\;\;\;\;\;\;1}\rangle_T=\frac14+\frac12\langle S^z_jS^z_{j+1}\rangle_T+\langle S^z_jS^z_{j+2}\rangle_T\nn\\
\langle S^x_jS^x_{j+2}\rangle_T &=&\frac12\langle e_{j\;1}^{\;\;\;2}\l(e_{j+1\;1}^{\;\;\;\;\;\;\;1}+e_{j+1\;2}^{\;\;\;\;\;\;\;2}\r)e_{j+2\;2}^{\;\;\;\;\;\;\;1}\rangle_T\nn.
\ee
The function $P(m)=\langle e_{1\;1}^{\;\;\;1}\ldots e_{m\;1}^{\;\;\;\;1}\rangle_T$ is referred to as the emptiness formation probability.

For the numerical treatment, it is convenient to reformulate equations \refeq{lna}, \refeq{g} such that the integrations are done along straight lines parallel to the real or imaginary axes. This procedure is due to Kl\"umper (cf. the book \cite{book} and references therein). We review the main steps here in order to introduce a notation which will prove to be convenient for our purposes. 

\begin{figure}
\begin{center}
\vspace{-1cm}
 \includegraphics[scale=0.8]{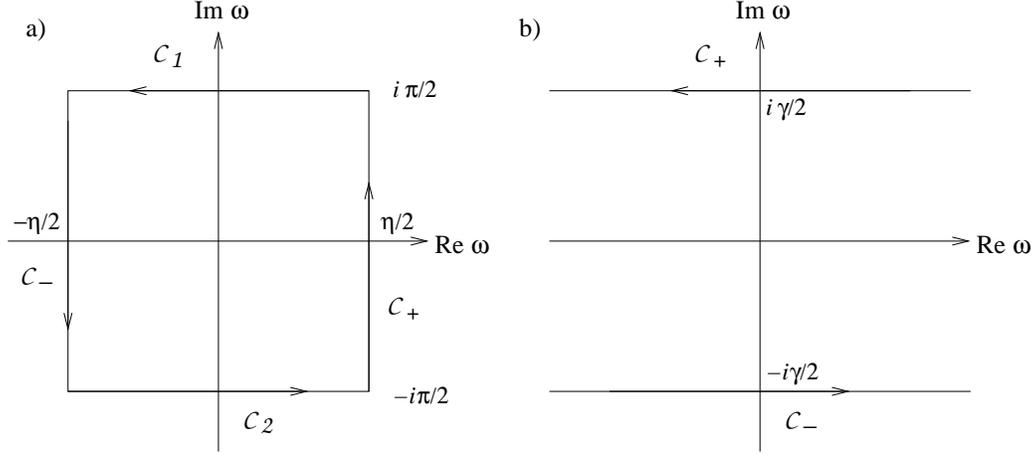}
\caption{Integration contours a) in the massive b) in the massless case.}
\label{fig1}
\end{center}
\end{figure}

The integration contour $\C$ is chosen according to fig. \ref{fig1}a) (\ref{fig1}b))  in the massive (massless) case. 
In the massive (massless) regime, Re $\C_\pm=\pm\eta/2\mp \e$ (Im $\C_\pm=\pm\g/2\mp \e$), where $\e$ is a small positive quantity; the limit $\e\to 0$ is included implicitly. According to eqs.\ \refeq{lna}, \refeq{g} the functions $\ln \a$, $G$ are given in the part of the complex plane enclosed by $\C$, once they have been calculated on $\C$. However, for the calculation of correlators, \refeq{gen}, these functions are only needed {\em on} $\C$. Since in the massive case, all functions are periodic with period $\rmi \pi$, integrals along $\C_{1,2}$ in opposite directions cancel each other. We thus have to calculate the auxiliary functions along two straight lines $\C_\pm$ parallel to the imaginary (real) axis in the massive (massless) case. 

Let us first focus on the calculation of the auxiliary functions in the massless case, the massive case is treated analogously afterwards. Subsequently, we comment on the calculation of the multiple integrals with the help of the auxiliary functions.  
%%%%%%%%%%%%%%%%%%%%%%%%%%%
%%% Massless case
%%%%%%%%%%%%%%%%%%%%%%%%%%
\subsection{Massless case}
We first concentrate on $0<\g\leq \pi/2$ and deal with the isotropic case $\g=0$ at the end of this section.
It is convenient to define
\begin{subequations}
\be
\b(x)&:=&\a(x+\rmi \g/2),\; \bb(x):=  \aa(x-\rmi\g/2)\label{lnb}\\
\B&:=& 1+\b, \;  \BB:= 1+\bb \label{lnbb}\; .
\ee
\end{subequations}
Note that $\bb=\b^*|_{h\to-h}$. We now perform a ''particle-hole-transformation'' in \refeq{lna} by substituting 
\be
\begin{array}{ll}
\ln(1+\a(x+\rmi \g/2))=\ln \B(x)& \mbox{on  } \C_+\\
\ln(1+\a(x-\rmi \g/2))=\ln \BB(x)+\ln \b(x-\rmi \g)& \mbox{on  } \C_- \; . 
\end{array}
\ee
Taking account of the simple pole of $\ln \b(x)$ at $x=-\rmi \g/2$, we arrive at
\be
\ln \b(x)&=&-\beta h+\frac{J \beta \sinh(\rmi \g)}{2}\,  \hat \k_{\rmi \g/2} (x)+\frac{1}{2\pi\rmi}\,\l[\hat \kappa_{\rmi \g}*\ln \B-\hat \k^{(+)}_{\rmi \g}*\ln \BB-\hat \k_{\rmi \g}*\ln \b\r](x)\label{lnb1}
\ee
with the definitions
\be
\hat \kappa_{\rmi \g}^{(+)}(x)&:=&\hat \kappa_{\rmi \g}(x+\rmi \g-\rmi \e)\nn\\
\l[f*g\r](x)&:=&\i f(x-y)g(y)\,\d y\label{conv1}\; .
\ee
The integrations in \refeq{lnb1} are now convolutions along the real axis, such that one can express $\ln \b$ in terms of $\ln \B$, $\ln \BB$ by solving an algebraic equation in Fourier space and transforming back to direct space. The result is
\begin{subequations}
\label{subb}
\be
\ln \b(x)&=& -\frac{\pi}{2(\pi-\g)}\,\beta h- \frac{J\beta\sin\g}{2} \,d(x) + \l[\k *\ln \B-\k_+*\ln \BB\,\r](x)\label{lnb2}\\
\ln \bb&=& \ln \b^*|_{h\to -h}\label{lnbb2}
\ee
\end{subequations}
with 
\be
d(x)&=&\frac{\pi}{\g\,\cosh\frac{\pi}{\g} x}\label{d1}\\
\kappa(x)&=& \i \frac{\sinh\l(\frac\pi2-\g\r)k\,\te^{\rmi k x}}{2\cosh\frac\g2k \, \sinh\l(\frac\pi2-\frac\g2\r)k}\, \frac{\d k}{2\pi}\nn\\
\kappa_+ (x)&=& \kappa(x+ \rmi \g- \rmi \e)\nn
\ee
Note that the shift by $\epsilon$ in \refeq{lnb2} is necessary to ensure integrability. In a similar fashion, eq.\ \refeq{g} is treated. Therefore we introduce the functions 
\be
G_\b(x):=G(x+\rmi \g/2,\xi)\,,\; G_\bb:=-G(x-\rmi \g/2,\xi)\label{defg},
\ee
where we suppressed the dependence on $\xi$ on the left-hand sides. It is now convenient to substitute
\be
\frac{G}{1+\a}&=&\l\{\begin{array}{ll} 
\displaystyle-\frac{G_\bb}{1+\bb^{-1}}& \mbox{on   } \C_-\\[0.4cm]
\displaystyle-\frac{G_\b}{1+\b^{-1}}+G_\b& \mbox{on   } \C_+
\end{array}\r.\label{tran1}\\
\frac{G}{1+\aa}&=&\l\{\begin{array}{ll} 
\displaystyle\frac{G_\bb}{1+\bb^{-1}}-G_\bb& \mbox{on   } \C_-\\[0.4cm]
\displaystyle\frac{G_\b}{1+\b^{-1}}& \mbox{on   } \C_+\,.
\end{array}\r.\label{tran2}
\ee
Then
\begin{subequations}
  \label{geq}
  \be
  G_\b(x)&=& \rmi\, d(x-\xi)+\l[\k *\frac{G_\b}{1+\b^{-1}}-\k_+*\frac{G_\bb}{1+\bb^{-1}}\r](x)\label{gb}\\
  G_\bb&=& G_\b^*|_{h\to -h}\label{gbb} .
  \ee
\end{subequations}
Using the substitution rules \refeq{tran1}, \refeq{tran2}, the general density matrix element \refeq{gen} is expressed in terms of $G_\b/(1+\b^{-1})$, $G_\bb/(1+\bb^{-1})$, $G_\b$, $G_\bb$. This expression is rather lengthy: For an $m$-point-function, one obtains a sum of $3^m$-many terms, each one an $m$-fold integral. Here we would like to spare the reader the general expression, but rather make the following comments:
\begin{itemize}
\item Eqs.\ \refeq{subb}, \refeq{gb} are solved numerically by iteration; the convolutions are done by applying the Fast-Fourier-Transform algorithm. Errors then arise from the truncation and the discretization of the integrals. 
\item  In \refeq{gen} derivatives of $G_{\b,\bb}$ with respect to $\xi$ enter. These can be calculated straightforwardly by taking the $n$-th derivative of \refeq{geq} with respect to $\xi$ which results in linear integral equations for $\partial^n_\xi G_{\b,\bb}$. 
\item The formulation in terms of the functions $G_{\b,\bb}$ is particularly suited for low temperatures. At $T=h=0$, $1/(1+\b^{-1})=1/(1+\bb^{-1})=0$ such that $G_{\b}=G_\bb$ is given by the inhomogeneity in \refeq{gb}. Furthermore, only one of the $3^m$-many terms in \refeq{gen} remains (namely the one containing only combinations of $G_\b, \,G_\bb$). Then in this term all functions are known. This is the well-known result of Jimbo et al. \cite{jm96} which has first been obtained for $\Delta>1$ in \cite{jim92} (for a proof in the critical case and an extension to $h\neq 0$ see also \cite{kit00}). 
\end{itemize}
The functions for the isotropic case $\g=0$ are obtained by rescaling $x\to \g x$, $G_{\b,\bb}\to \g G_{\b,\bb}$. Then in \refeq{gen} algebraic functions occur instead of hyperbolic ones, and the integration kernel $\kappa$ takes the form
\be
\kappa(x)&=& \i \frac{\te^{\rmi k x}}{1+\te^{|k|}} \, \frac{\d k}{2\pi}\nn.
\ee
%%%%%%%%%%%%%%%%%%%%%%%%%%%%%
%%%% Massive case
%%%%%%%%%%%%%%%%%%%%%%%%%%%% 
\subsection{Massive case}
Let us go through the changes which have to be done in the massive case with respect to the massless case. Analogously to \refeq{lnb}, \refeq{lnbb}, we define the functions
\be
 \b(x)&=& \a(\rmi x+\eta/2),\; \bb(x)= \aa(\rmi x-\eta/2)\nn\\
\B&:=&1+\b, \; \BB:= 1+\bb\nn\; .
\ee
We use the same symbols $\b,\bb$ etc. as in the massless case. Note that here $-\pi/2\leq x \leq \pi/2$ and the $\b$-functions are periodic on the real axis with period $\pi$. Accounting for the pole of $\ln \b(x)$ at $x=\rmi \eta/2$ we find the following equations 
\be
\ln \b(x)&=& -\beta h-\frac{J \beta\sinh\eta}{2} \, \hat \kappa_{\eta/2}(x) +\frac{1}{2\pi}\l[\hat \k_\eta *\ln \B-\hat \k_\eta *\ln \b-\hat \k^{(-)}_\eta *\ln \BB\,\r](x)\label{lnbm}\\
\ln \bb(x)&=&\l.\ln \b^*\r|_{h\to-h}\nn,
\ee
where the kernel and the convolutions are now defined as
\be
\hat\k_\eta(x)&=&\frac{\sinh(2\eta)}{\sin(x+\rmi \eta)\, \sin(x-\rmi \eta)}\nn\\
\l[f*g\r](x)&=&\int_{-\pi/2}^{\pi/2} f(x-y)g(y)\, \d y\label{conv2}.
\ee
Because of periodicity, eq.\ \refeq{lnbm} is manipulated further in Fourier space after performing a discrete Fourier transformation. Upon transforming back one obtains
\be
\ln \b(x)&=& -\frac{\beta h}{2}-\frac{J\beta\sinh\eta}{2}\,d(x) +\l[\kappa *\ln \B-\kappa_-*\ln\BB\,\r](x)\label{lnbm2}\\
d(x)&=&\frac{2K}{\pi}\dn \l(\frac{2Kx}{\pi},\rmi\frac{\eta}{\pi}\r)\label{d2}\\
\kappa(x)&:=& \sum_{n=-\infty}^\infty \frac{\te^{\rmi 2n x}}{\pi\l(1+\te^{2  \eta|n|}\r)}\,\nn\\
\kappa_- (x)&:=& \kappa(x- \rmi \eta+ \rmi \epsilon).
\ee
The driving term \refeq{d2} is written in terms of the Jacobi elliptic function dn$(x,\tau)$ \cite{ww}, with Fourier series expansion
\be
\dn(x,\tau)&=& \frac{\pi}{2K(\tau)}\sum_{n=-\infty}^{\infty} \, \frac{\te^{\rmi \pi n x/K(\tau)}}{\cos n\pi \tau},\label{defdn}
\ee
in the strip $\l|\Im (x)\r|<\Im(\tau\,K(\tau))$ of the complex plane. In \refeq{defdn} the constant $K(\tau)$ is defined through $K(\tau):=\frac{\pi}{2} \vartheta_3^2(0,\tau)$
and $\vartheta_3$ is one of Jacobi's Theta-functions \cite{ww}. In very close analogy one derives equations for $G_{\b,\bb}$, defined by
\be
G_\b(x)&:=& G(\rmi x+\eta/2,\xi)\,,\;G_\bb(x):= -G(\rmi x-\eta/2,\xi) \nn\; ,
\ee
where, as in the massless case, the $\xi$-dependence is not noted explicitly in $G_{\b,\bb}$.
The symbolical substitutions \refeq{tran1}, \refeq{tran2} still hold, with $\C_{\pm}$ defined in fig.\ \ref{fig1}a), so that
\be
G_\b(x)&=&-d(x-\xi)+\l[\kappa *\frac{G_\b}{1+\b^{-1}}-\kappa_-*\frac{G_\bb}{1+\bb^{-1}}\r](x)\label{gm}\\
G_\bb&=& G_\b^*|_{h\to -h}\label{gmbb} .
\ee
%%%%%%%%%%%%%%%%%%%%%%
%%%Calculating the integrals
%%%%%%%%%%%%%%%%%%%%%%
\subsection{Calculating the integrals}
Let us consider eq.\ \refeq{gen}, where the substitutions implied by \refeq{tran1}, \refeq{tran2} have been performed. One observes that due to the product of hyperbolic functions in the denominator in eq.\ \refeq{gen}, multiple poles may occur. We explain in the following how to treat these poles numerically and show that contributions due to these poles cancel each other if there is more than one pole. For definiteness we treat the massless case, however, the arguments are directly transferable to the massive case. 

The product $\prod_{1\leq a<b\leq m} \sinh(\om_a-\om_b-\rmi \g)$ with 
\be
\om_\al=\l\{\begin{array}{ll} 
x_\al+\rmi \g/2 & \mbox{on   } \C_+\\
x_\al-\rmi \g/2 & \mbox{on   } \C_-
\end{array}\r.
\ee
acquires a zero if there are $j,k$ such that $\om_j=x_j+\rmi \g/2-\rmi \e$, $\om_k=x_k-\rmi \g/2+\rmi \e$. This leads to a factor
\be
\frac{1}{\sinh(x_j-x_k-\rmi \e)}&=& \text{P}\l[\frac{1}{\sinh(x_j-x_k)}\r]+\rmi \pi \delta(x_j-x_k)\label{pol1}\; .
\ee
When taking the principal value numerically, one has to account for the regular part at $x_j=x_k$, defined by $f_{reg}(x_j,x_j)=0$: 
\be
\text{P} \i \frac{f(x_j,x_k)}{\sinh(x_j-x_k)}\, \d x_k&=&\int_{-\infty}^{x_j-\e}+\int_{x_j+\e}^\infty \frac{f(x_j,x_k)}{\sinh(x_j-x_k)}\,\d x_k -2\epsilon \partial_{x_k} f_{reg}(x_j,x_k)|_{x_k=x_j}\, \label{pol2}.
\ee
The last terms in eqs.\ \refeq{pol1}, \refeq{pol2} are additional terms caused by the pole. 

Now consider the case of multiple poles. Set $\om_j=x_j+\rmi \g/2$ in the product $\prod_{1\leq a<b\leq m} \sinh(\om_a-\om_b-\rmi \g)$ and all other $\om_k=x_k-\rmi \g/2$. Then if $j=m-1$, there is one simple pole. If $j<m-1$, the product $\sinh(x_j-x_{m-1}-\rmi \e)\sinh(x_j-x_m-\rmi \e)$ occurs in the denominator. It is not difficult to show that the additional contributions due to these two poles either vanish directly  or cancel each other. One should be aware of the fact that the $\delta$-function in \refeq{pol1} yields no contribution if $f\equiv f_{reg}$. This is the case for $T=0$, where all functions are real and the determinant in \refeq{gen} vanishes if any two arguments of the auxiliary functions are equal. 

We now address the question of possible simplifications of the multiple integral representation \refeq{gen}. Obviously, in the case $m=2$ the double integration has to be equal to one single integration, since the nearest-neighbor correlators are obtained by taking the derivative of the free energy per lattice site $f$ with respect to the anisotropy $\Delta$, and the free energy is given by a single integral according to \cite{kl93}. Especially
\begin{subequations}
\be
\langle S_j^zS_{j+1}^z\rangle_T&=& \frac14+\frac{\partial_\Delta f }{J} \label{nn1}\\
\langle S_j^xS_{j+1}^x\rangle_T&=& \frac{u}{2J}  -\frac{\Delta}{2} \langle S_j^zS_{j+1}^z-1/4\rangle_T\label{nn2}\\
\langle S_j^zS_{j+1}^z\rangle_T&=&\langle S_j^xS_{j+1}^x\rangle_T=\frac{1}{12}+\frac{u}{3 J},\qquad \Delta=1\label{nn3}, 
\ee
\end{subequations}
where $u=\partial_\beta(\beta f)$ is the inner energy per lattice site and $f$ is given by
\be
f= e_0-\frac{T}{2\pi}\l[d*\ln\B\BB\r](0). \nn 
\ee
Here $d(x)$ is given by eq.\ \refeq{d1} (eq.\ \refeq{d2}) in the massless (massive) case; the convolution for both cases is defined in eq.\ \refeq{conv1} (eq.\ \refeq{conv2}). The ground state energy is given by
\be
e_0&=&\l\{ 
\begin{array}{ll}
\displaystyle -\frac{J\,\sin\g}{2}\,\i \frac{\sinh\l(\frac\pi2-\frac\g2\r) k}{2\cosh\frac{\g k}{2} \, \sinh \frac{\pi k}{ 2}}\, \d k& \mbox{massless}\\[0.4cm]
\displaystyle -\frac{J \sinh\eta}{2} \sum_{n=-\infty}^\infty \frac{\te^{-\eta |n|}}{\cosh\eta n}& \mbox{massive.}
\end{array}\r.\nn
\ee
Eqs.\ \refeq{nn1}, \refeq{nn2} also allow us to extract the leading order in $T$ at low temperatures. We consider here the case of vanishing magnetic field, $h=0$. From \cite{klpei}, using dilogarithms, one finds for $T\to 0$ in the massless case
\be
f=e_0-T^2\frac{\g}{3J\sin \g}\nn.
\ee
Thus for $T\to 0$
\begin{subequations}
\be
\langle S^z_jS^z_{j+1}\rangle_T&=& \langle S^z_jS^z_{j+1}\rangle_{T=0} +(T/J)^2 \,\frac{\l(1-\g\,\cot\g\r)}{3 \sin^2 \g}\label{aatt}\\
\langle S^x_jS^x_{j+1}\rangle_T&=& \langle S^x_jS^x_{j+1}\rangle_{T=0} +(T/J)^2\,\frac{\l(\g-\cot\g+\g\,\cot^2\g\r)}{6 \sin\g}\label{xxtt}\\
\langle S^z_jS^z_{j+1}\rangle_T&=&\langle S^x_jS^x_{j+1}\rangle_T=\langle S^z_jS^z_{j+1}\rangle_{T=0}+\frac{T^2}{9 J^2},\; \Delta=1\label{aattiso}.
\ee
\end{subequations}
The correlator $\langle S^z_jS^z_{j+1}\rangle_{T=0}$ as a function of $\Delta$ is plotted in fig.\ 3 of ref.\ \cite{joh73}. This result has been extended to all temperatures in the range $-1\leq \Delta\leq 1$ in \cite{fab98}, fig.\ 2. Note from \refeq{aatt}-\refeq{aattiso} that the sign of the $T^2$-coefficient is positive, so that the antiferromagnetic correlations first decrease with increasing temperature. 
%Numerical values from exact diagonalization are given for $\langle S^z_jS^z_{j+m-1}$ for $m=2,\ldots,10$ with anisotropy $-1<\Delta<0$ in \cite{fab99}. 

In the massive case, for finite $\eta$, the leading temperature-dependent order may be obtained by a saddle-point integration, very similar to \cite{joh72}. Then
\be
f&=&-e_0-A^{1/2} T^{3/2} \te^{-B/T}\nn\\
A&=&\frac{k'}{2J  k^2 K\sinh\eta}\nn\\
B&=& J\frac{k'\,K}{\pi} \sinh\eta\nn
\ee
with the moduli $k^{1/2}:=\vartheta_2(0,\rmi\eta/\pi)/\vartheta_3(0,\rmi\eta/\pi)$ and $\l(k'\r)^{1/2}:=\vartheta_4(0,\rmi\eta/\pi)/\vartheta_3(0,\rmi\eta/\pi)$. It is now not difficult to show that 
\be
\langle S^z_jS^z_{j+1}\rangle_T&=& \langle S^z_jS^z_{j+1}\rangle_{T=0} +C T^{1/2}\te^{-B/T} \nn\\
\langle S^x_jS^x_{j+1}\rangle_T&=& \langle S^x_jS^x_{j+1}\rangle_{T=0} +D T^{1/2} \te^{-B/T}\nn\\
C&=& A^{1/2} \frac{k'K}{\pi}\l(\coth\eta+\sum_{n=1}^\infty\frac{1}{\cosh^2\eta(n-1/2)}\r)>0\nn\\
D&=& \frac{1}{2J}\l(A^{1/2} B-J\, C \cosh\eta\r)<0\nn\; .
\ee
At first sight, the last inequality is surprising: The antiferromagnetic correlations of $\langle S^x_jS^x_{j+1}\rangle_T$ increase with increasing temperature. Since $\lim_{T\to\infty}\langle S^x_jS^x_{j+1}\rangle_T=0$, we expect a minimum at finite temperature. This will be confirmed in the next section. Intuitively, this result can be understood as follows: At $T=0$, the Ising-like anisotropy leads to a Ne\'el-like ground state, which enhances the alignment of the spins along the $z$-direction. An initial increase of the temperature reduces $\l|\langle S^z_jS^z_{j+1}\rangle_T\r|$, so that quantum fluctuations of the spins are favored, and $\l|\langle S^x_jS^x_{j+1}\rangle_T\r|$ increases. At high enough temperatures the spins become uncorrelated, so that both correlation functions decay to zero.

As far as the low-temperature behavior of correlation functions in the massive regime is concerned, we would like to note that using the saddle-point integration technique, one finds that
\be
\langle e_{1\;\beta_1}^{\;\;\;\al_1}\ldots e_{m\;\beta_m}^{\;\;\;\al_m}\rangle_T&=&\langle e_{1\;\beta_1}^{\;\;\;\al_1}\ldots e_{m\;\beta_m}^{\;\;\;\al_m}\rangle_{T=0}+C_m T^{1/2}\te^{-B/T}+\Or\l(T^{3/2}\te^{-B/T}, T\te^{-2B/T}\r)\nn,
\ee
for arbitrary $m$, with some coefficient $C_m$.

Up to now, we were not able to derive eqs.\ \refeq{nn1}, \refeq{nn2} from \refeq{gen}. Still, the latter equations provide a useful test for the accuracy of our numerical integration scheme for $m=2$, as explained in more detail in the following section. For $m>2$, simplifications of \refeq{gen} have not been performed yet for any finite temperatures except at $\Delta=0$ and $ \Delta\to\infty$, see below. However, for $T=0$, where the integrand is known explicitly, the multiple integrals have been simplified considerably in the isotropic regime \cite{boo01,boo03,boo05}. In the anisotropic regime, the multiple integrals have been reduced to single integrals for $m\leq 4$ \cite{kat03,kat04}. These works allow us to estimate the numerical accuracy at $T=0$. In the other extreme, at $T=\infty$, the spins are uncorrelated so that in this limiting case we also have explicit results which can be checked. Note, however, that it is rather non-trivial to reproduce these high-temperature results from the general formula \refeq{gen}. This was done only very recently by Tsuboi and Shiroishi in \cite{tsu05}, where a high-temperature expansion of $P(m)$ has been performed to high order in $1/T$ for the isotropic case. This is one further result which will serve as a reference point for estimating the accuracy of our numerics for $m=3$.   

The cases $\Delta=0$ (free fermions) and $\Delta\to\infty$ such that $J\cdot \Delta=:c$ finite (Ising limit) allow for substantial simplifications. Details will be explained in a separate publication \cite{gs05}, here we only give the results. For $\Delta=0$, that is $\g=\pi/2$, the convolutions in \refeq{lnb2}, \refeq{gb} disappear, so that all functions are known explicitly for all temperatures. Then
\be
P(m)=\det_m\l[\int_{-\pi}^\pi \frac{\d p}{2\pi} \frac{\te^{\rmi(a-b)p}}{1+\exp\l(\beta J\cos p-\beta h\r)}\r]_{ab},\nn
\ee
in agreement with \cite{shi01}. 
In the other extreme, the Ising limit, the driving term and integration kernels become constant ($d(x)\to 1$, $\kappa(x)\to1/(2\pi)$) such that the auxiliary functions do not depend on $x$ and the integrations are trivial. One finds \cite{bax82,suz02,gs05}
\be
P(m)&=&\frac{1+\langle\s^z\rangle_T}{2}\l[\frac{1+\langle\s^z\rangle_T}{2}+\frac{1-\langle\s^z\rangle_T}{2}\frac{\la_2}{\la_1}\r]^{m-1}\label{is}\\
\la_{1,2}&=&\te^{-\beta c/4} \l(\cosh\frac{\beta h}{2}\pm \sqrt{\sinh^2\frac{\beta h}{2}+\te^{\beta c}}\r)\nn\\
\langle\s^z\rangle_T&=&\frac{\sinh\frac{\beta h}{2}}{\sqrt{\sinh^2\frac{\beta h}{2}+\te^{\beta c}}}\nn .
\ee

%%%%%%%%%%%%%%%%%%%%%%%%%%%
%%% Results
%%%%%%%%%%%%%%%%%%%%%%%%%%
\section{Results}
We first present results of the next-neighbor correlation functions, comparing data from the double integration with those from the single integration. Results for next-nearest neighbors will be given subsequently. Correlation functions for longer distances can be evaluated in the same way, however, the numerical cost increases exponentially with the number of integrations. That is why for the time being in this feasibility study we do not go beyond $m=3$. 

%%%%%%%%%%%%%%%%%%%%%
%% Nearest Neighbors
%%%%%%%%%%%%%%%%%%%%%
\subsection{Nearest neighbors}
\begin{figure}
\begin{center}
\vspace{-2cm}
 \includegraphics[angle=-90, scale=0.5]{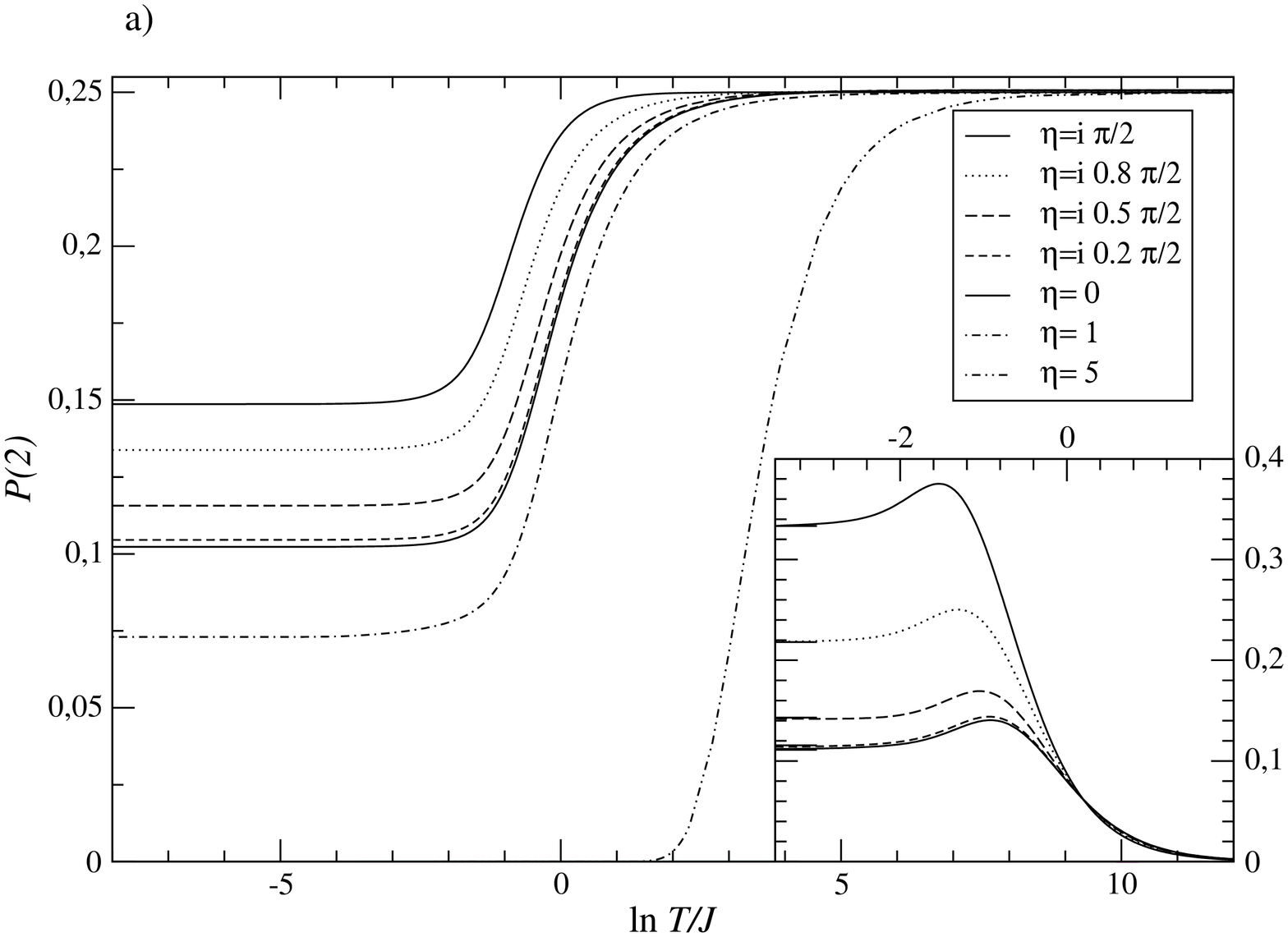}
\vspace{-1cm}
\includegraphics[angle=-90, scale=0.5]{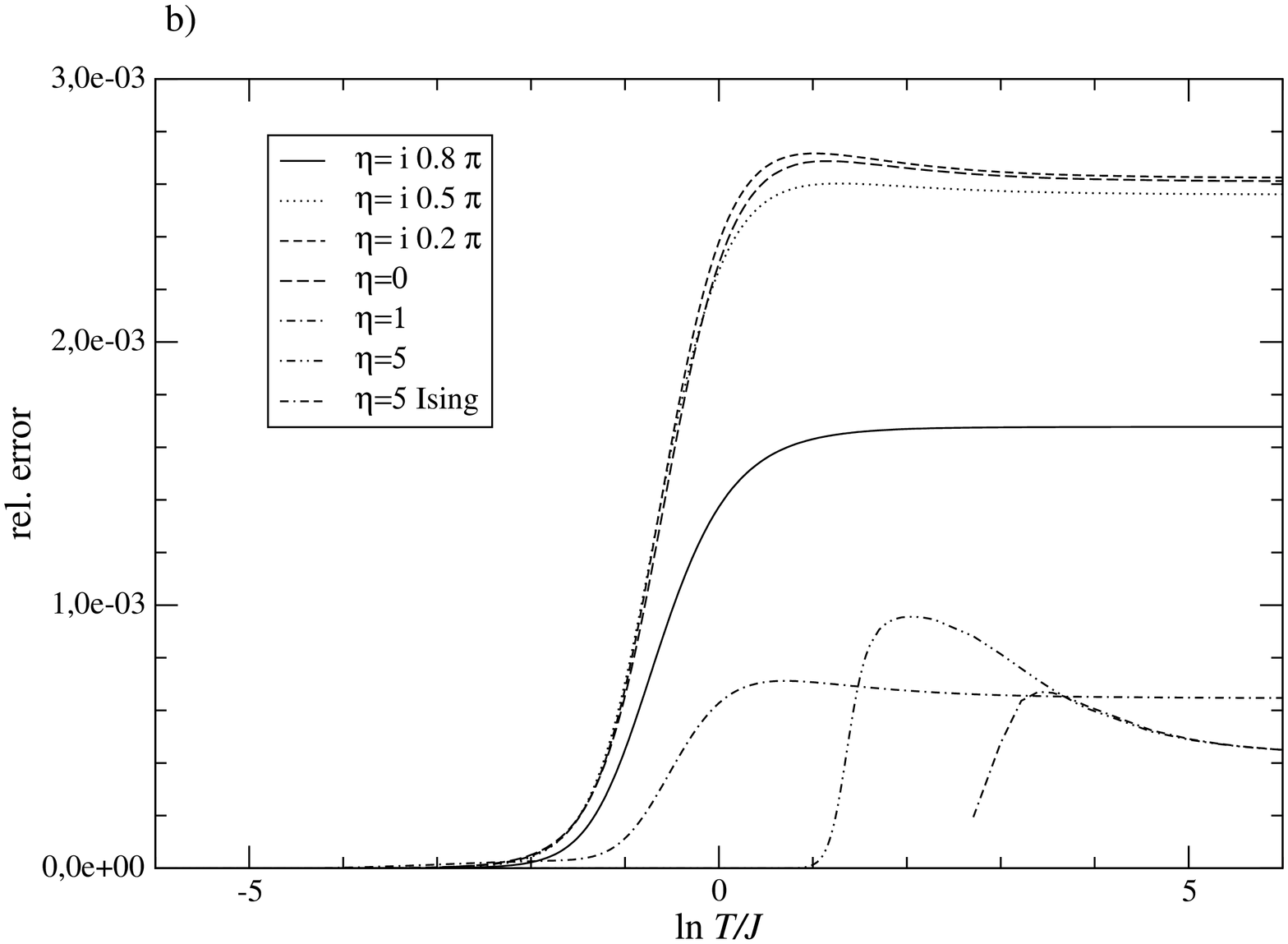}
\caption{a) Emptiness formation probability $P(2)$ for nearest neighbors; the upper curve corresponds to the $XX$ limit $\eta=\rmi \pi/2$. The inset shows the $T^2$-coefficient at low temperatures in the massless case. Straight lines indicate the exact values according to eqs.\ \refeq{aatt}, \refeq{aattiso}. b) Relative error of $P(2)$. The curve labeled by ``$\eta =5$ Ising'' shows the relative difference between the $\eta=5$-values of a) and the Ising limit eq.\ \refeq{is}.}
\label{fig2}
\end{center}
\end{figure}

\begin{figure}
\begin{center}
%\vspace{-1cm}
 \includegraphics[angle=-90, scale=0.5]{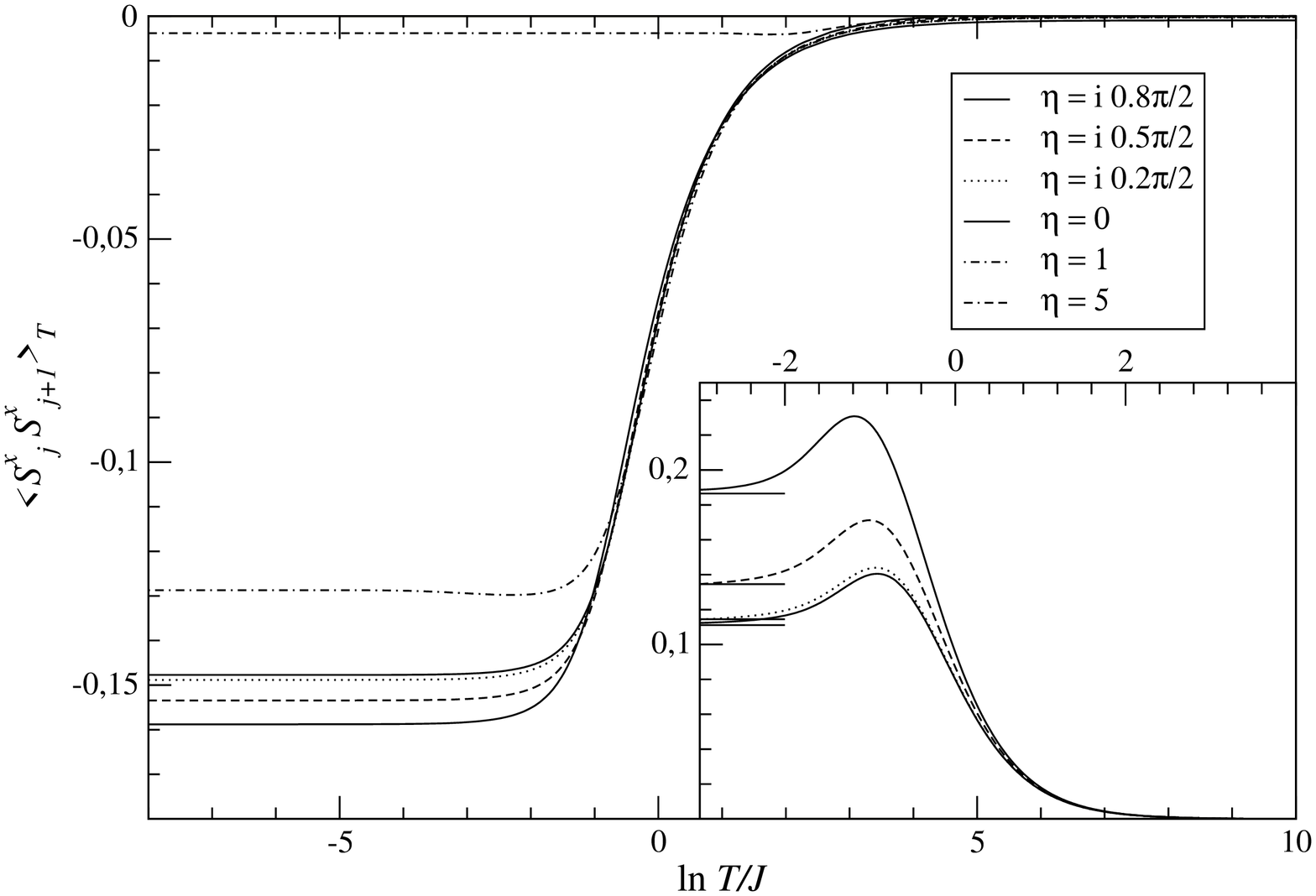}
%\caption{}
\caption{Correlation function $\langle S^x_jS^x_{j+1}\rangle$ for nearest neighbors; the lower curve corresponds to $\eta=\rmi 0.8\pi/2$. The inset shows the $T^2$-coefficient at low temperatures in the massless case. Straight lines indicate the exact values according to eq.\ \refeq{xxtt}.}
\label{fig3}
\end{center}
\end{figure}

Figs.\ \ref{fig2}, \ref{fig3} show $P(2)$, $\langle S^x_jS^x_{j+1}\rangle_T$, as calculated from the double integral, for $P(2)$ together with the relative error with respect to the data obtained from the derivative of the free energy according to \refeq{nn1}, \refeq{nn2}. The single integration results can be considered as exact; the numerical error is less than 8 digits for all temperatures. We checked that the relative error for $\langle S^x_jS^x_{j+1}\rangle_T$ vanishes at low temperatures as for $P(2)$; at high temperatures, the absolute error is about $10^{-4}$. 

In both cases, the error increases with increasing temperature, reaching maximal values at $T\approx J$ near the isotropic point. Generally, the error is considerably larger at high than at low temperatures. This is due to two reasons:
\begin{itemize}
\item At finite temperature, the auxiliary functions are given numerically from the solutions of the integral equations, whereas at $T=0$, they are known explicitly. The numerical iteration scheme necessarily induces errors due to the truncation and discretization of the integrals.
\item The additional contribution from poles in eq.\ \refeq{pol1} increases with increasing temperature and causes an additional error in the integrations.
\end{itemize}
We have verified that the error is reduced by decreasing the distance between two successive integration points. We conclude that the errors are mainly due to the truncation and discretization of the integrals and can be made arbitrarily small at the expense of increasing computation time.

Considering the low-temperature behavior of $\langle S^x_jS^x_{j+1}\rangle_T$, one finds an initial decrease from $T=0$ with a minimum at $T_0$, this is illustrated in fig.\ \ref{figmass}.  We commented on this phenomenon in the previous section. From fig.\ \ref{figmass} one observes that in the Ising limit, the maximum of $\langle S^x_jS^x_{j+1}\rangle_T/\langle S^x_jS^x_{j+1}\rangle_{T=0}$ is located at $T_0 /(\Delta J)\approx 0.168$.

%%%%%%%%%%%%%%%%%%%%%
%% Next-Nearest Neighbors
%%%%%%%%%%%%%%%%%%%%%
\subsection{Next-Nearest neighbors}
We calculated next-nearest neighbor correlation functions for the isotropic and the massive case. In the massless case, the auxiliary functions in \refeq{gen} are multiplied by a kernel which, depending on the integration variables, can increase exponentially, so that the auxiliary functions have to be determined extremely accurately over the whole integration range. This problem is still controllable for nearest neighbors, but becomes severe for $m>2$. In the isotropic case, however, one deals with algebraic functions, which are much better to handle numerically. At anisotropy parameter $\Delta>1$ the integration range is finite, so that those problems do not occur.

\begin{figure}
\begin{center}
\vspace{-2cm}
 \includegraphics[angle=-90, scale=0.5]{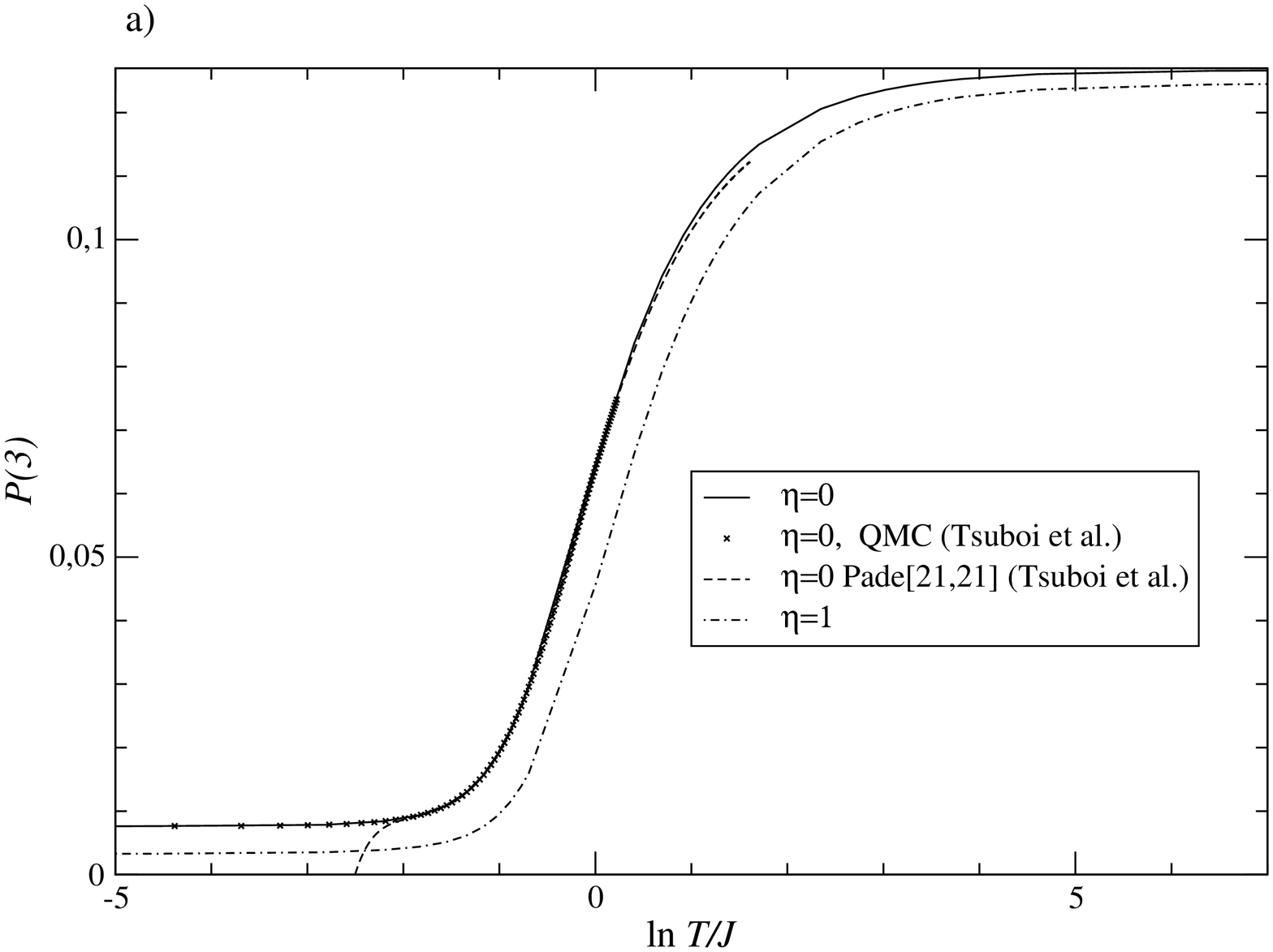}
\vspace{-1cm}
 \includegraphics[angle=-90, scale=0.5]{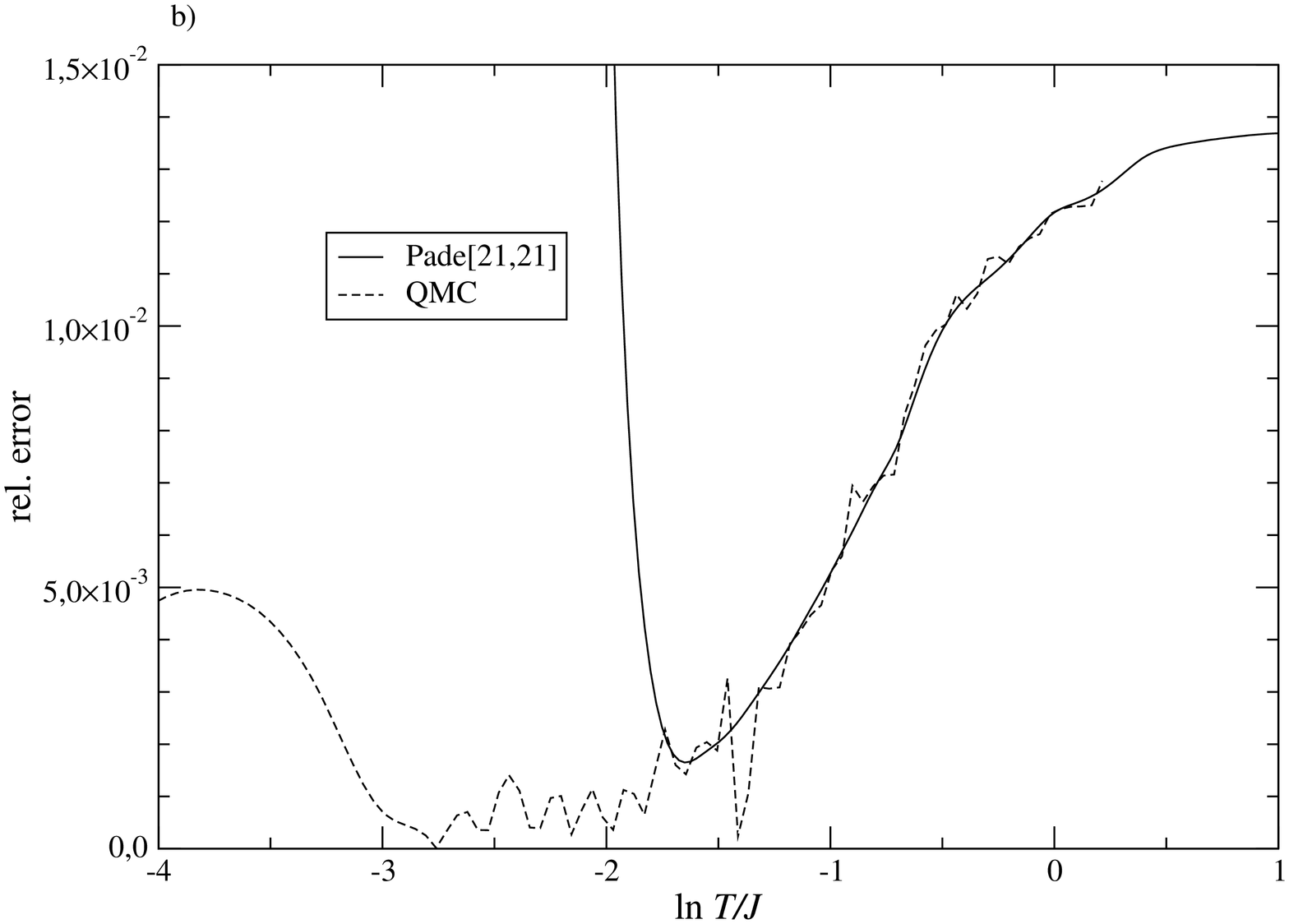}
\caption{a) Emptiness formation probability $P(3)$ for next-nearest neighbors. Shown are results from the 3-fold integral, from QMC \cite{tsu05} and high-temperature expansion (HTE) combined with Pad\'e approximation \cite{tsu05}. b) Relative error of the data from the 3-fold integral with respect to the QMC and the HTE-data.}
\label{fig4}
\end{center}
\end{figure}

In fig.\ \ref{fig4} we show the emptiness formation probability $P(3)$. In the isotropic case, we compare our data with the high-temperature expansion of \cite{tsu05} and with Quantum Monte Carlo (QMC) data which were obtained in \cite{tsu05}. The relative error of our data shown in fig.\ \ref{fig4} is larger than in fig.\ \ref{fig2}, which is due to the fact that the data in \refeq{fig4} were calculated by using only half of the number of integration points compared to the $m=2$ case. We checked at single temperature values that the error is of the same order as in the $m=2$ case when the calculations are done with the same numerical accuracy. However, the computational costs would be unreasonably high. At lowest temperatures, a comparison with \cite{boo01,kat03,kat04} yields relative errors of the order $2\cdot10^{-4}$. 

\begin{figure}
\begin{center}
%\vspace{-1cm}
 \includegraphics[angle=-90, scale=0.5]{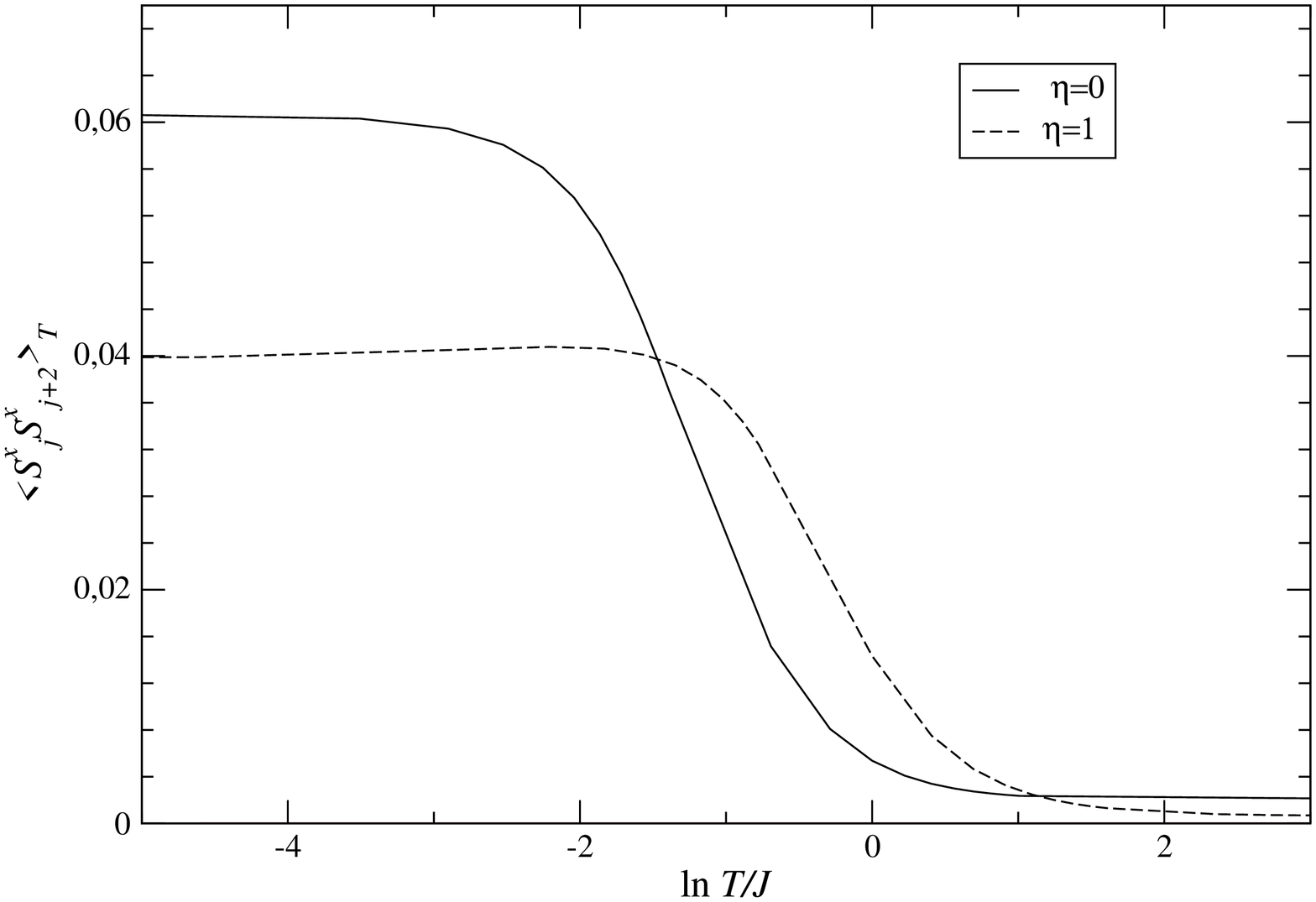}
\caption{The correlation function $\langle S^x_jS^x_{j+2}\rangle_T$ for $\eta=0,1$. Note that at low temperatures for $\Delta>1$, the correlation increases, before it starts to decrease to zero.}
\label{fig5}
\end{center}
\end{figure}

\begin{figure}
\begin{center}
%\vspace{-1cm}
 \includegraphics[angle=-90, scale=0.5]{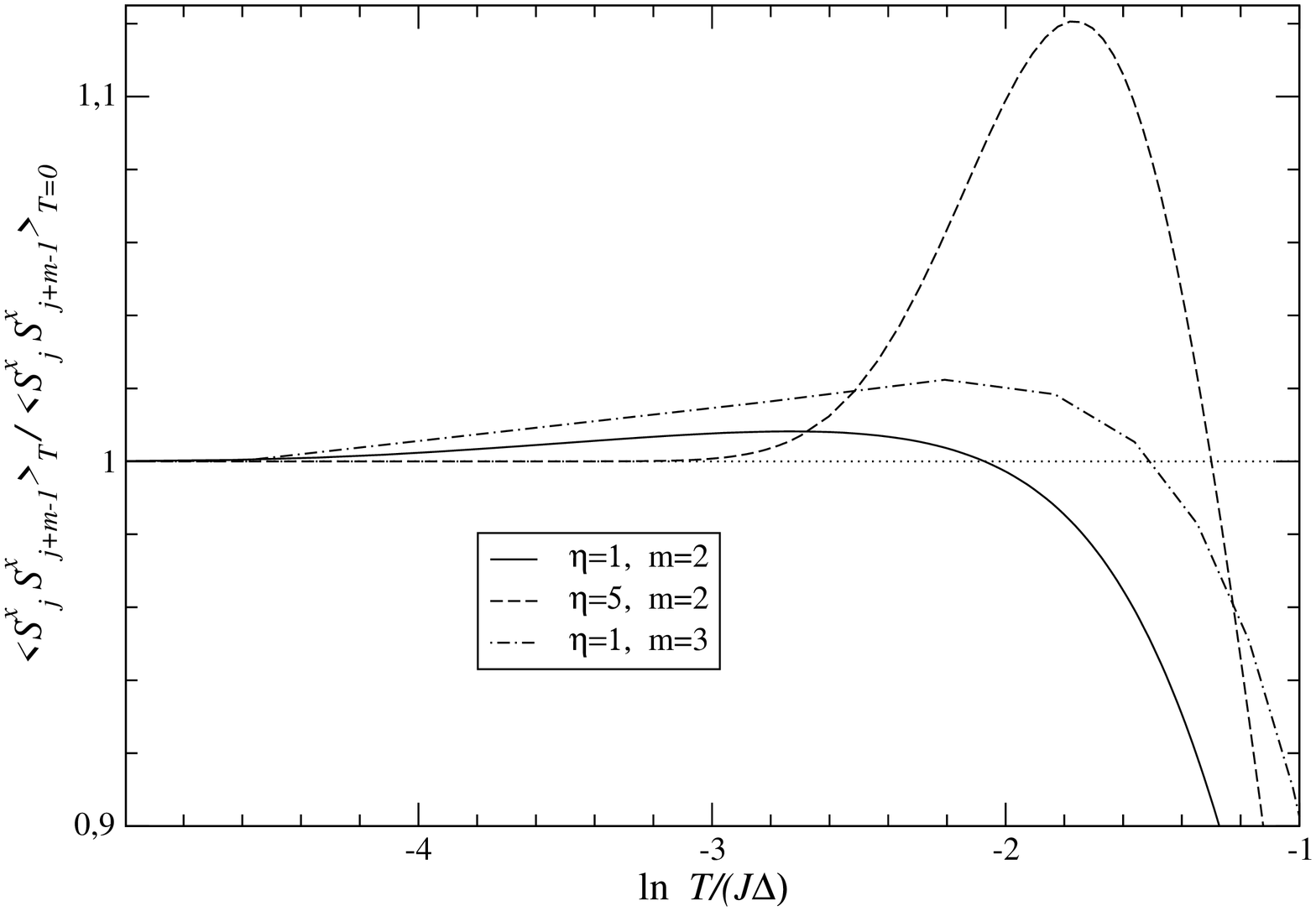}
\caption{The correlation function $\langle S^x_jS^x_{j+m-1}\rangle_T$ divided by its zero-temperature value $\langle S^x_jS^x_{j+m-1}\rangle_{T=0}$ for $m=2,3$ in the massive regime $\Delta>1$. The dotted horizontal line gives the $T=0$-value 1. The temperature is normalized with respect to $J\Delta$.}
\label{figmass}
\end{center}
\end{figure}

As a second example, $\langle S^x_{j}S^x_{j+2}\rangle_T$ is depicted in fig.\ \ref{fig5}. At lowest temperatures we again compare our data with the explicit results of \cite{kat03,kat04} which yields a relative error of $7\cdot 10^{-3}$. At highest temperatures, $\langle S^x_{j}S^x_{j+2}\rangle_T=0$. The absolute error here is of the order $10^{-3}$. As in the $m=2$-case we observe an increase of the antiferromagnetic correlation with increasing $T$, reaching a maximum at some finite temperature $T_0$ in the massive regime. Fig.\ \ref{figmass} illustrates this behaviour, where $\langle S^x_jS^x_{j-m+1}\rangle_T/\langle S^x_jS^x_{j-m+1}\rangle_{T=0}$ is depicted for $m=2,3$ and $\Delta>1$.  
%%%%%%%%%%%%%%%%%%%%%%%%%
%%% Summary and Outlook
%%%%%%%%%%%%%%%%%%%%%%%%
\section{Summary and Outlook}
We demonstrated that an accurate numerical evaluation of the formula \refeq{gen} is possible by considering the special cases $m=2$ with $\Delta \geq 0$ and $m=3$ with $\Delta\geq 1$. The numerical error was estimated precisely by comparing with alternative approaches. We have found that this error is only due to the discretization and, for $\eta =\rmi \g$, the truncation of the integrals. For $m=3$ in the massless case we encountered problems hard to deal with numerically. Also the computational costs of an extension of the numerics to $m>3$ become very high if one wants to maintain the accuracy. Thus, the next step should be to address the question of in how far the multiple integrals can be simplified. Here we would like to mention two approaches. 

In \cite{tsu05} an high-temperature expansion (HTE) has been applied to calculate $P(m)$. The excellent agreement of these data with those obtained by numerically evaluating the multiple integrals directly (cf. fig.\ \ref{fig4}), even at temperatures below the crossover region highlights the applicability of the HTE method. On the other hand, in the works \cite{boo01,boo04,kat03,kat04} the multiple integrals have been reduced to single integrals at $T=0$. The case $m=2$ shows that simplifications are possible also at finite $T$ (compare also \cite{gks2}), where the integrand in \refeq{gen} is given implicitly by integral equations. Further progress may be expected from extending the recent results \cite{boo04,boo04b} for the inhomogeneous generalizations of the correlation functions of the $XXZ$ chain to finite $T$. This issue is the subject of current research.
  
\section*{Acknowledgments}
We are very grateful to M. Shiroishi and Z. Tsuboi for sending us their QMC- and HTE-data. We would also like to acknowledge very helpful and stimulating discussions with A. Kl\"umper and A. Seel.


\newpage


\begin{thebibliography}{33}
\expandafter\ifx\csname natexlab\endcsname\relax\def\natexlab#1{#1}\fi
\expandafter\ifx\csname bibnamefont\endcsname\relax
  \def\bibnamefont#1{#1}\fi
\expandafter\ifx\csname bibfnamefont\endcsname\relax
  \def\bibfnamefont#1{#1}\fi
\expandafter\ifx\csname citenamefont\endcsname\relax
  \def\citenamefont#1{#1}\fi
\expandafter\ifx\csname url\endcsname\relax
  \def\url#1{\texttt{#1}}\fi
\expandafter\ifx\csname urlprefix\endcsname\relax\def\urlprefix{URL }\fi
\providecommand{\bibinfo}[2]{#2}
\providecommand{\eprint}[2][]{\url{#2}}

\bibitem[{\citenamefont{G\"ohmann
  et~al.}(2004{\natexlab{a}})\citenamefont{G\"ohmann, Kl\"umper, and
  Seel}}]{gks1}
\bibinfo{author}{\bibfnamefont{F.}~\bibnamefont{G\"ohmann}},
  \bibinfo{author}{\bibfnamefont{A.}~\bibnamefont{Kl\"umper}},
  \bibnamefont{and} \bibinfo{author}{\bibfnamefont{A.}~\bibnamefont{Seel}},
  \bibinfo{journal}{J. Phys. A} \textbf{\bibinfo{volume}{37}},
  \bibinfo{pages}{7625} (\bibinfo{year}{2004}{\natexlab{a}}).

\bibitem[{\citenamefont{G\"ohmann
  et~al.}(2004{\natexlab{b}})\citenamefont{G\"ohmann, Kl\"umper, and
  Seel}}]{gks2}
\bibinfo{author}{\bibfnamefont{F.}~\bibnamefont{G\"ohmann}},
  \bibinfo{author}{\bibfnamefont{A.}~\bibnamefont{Kl\"umper}},
  \bibnamefont{and} \bibinfo{author}{\bibfnamefont{A.}~\bibnamefont{Seel}},
  \bibinfo{journal}{Physica B} \textbf{\bibinfo{volume}{807}}, \bibinfo{pages}{359} (\bibinfo{year}{2005}{\natexlab{b}}).

\bibitem[{\citenamefont{G\"ohmann et~al.}(2005)\citenamefont{G\"ohmann,
  Kl\"umper, and Seel}}]{gks3}
\bibinfo{author}{\bibfnamefont{F.}~\bibnamefont{G\"ohmann}},
  \bibinfo{author}{\bibfnamefont{A.}~\bibnamefont{Kl\"umper}},
  \bibnamefont{and} \bibinfo{author}{\bibfnamefont{A.}~\bibnamefont{Seel}},
  \bibinfo{journal}{J. Phys. A} \textbf{\bibinfo{volume}{38}},
  \bibinfo{pages}{1833} (\bibinfo{year}{2005}).

\bibitem[{\citenamefont{Suzuki}(1985)}]{suz85}
\bibinfo{author}{\bibfnamefont{M.}~\bibnamefont{Suzuki}},
  \bibinfo{journal}{Phys. Rev. B} \textbf{\bibinfo{volume}{31}},
  \bibinfo{pages}{2957} (\bibinfo{year}{1985}).

\bibitem[{\citenamefont{Suzuki and Inoue}(1987)}]{suz87}
\bibinfo{author}{\bibfnamefont{M.}~\bibnamefont{Suzuki}} \bibnamefont{and}
  \bibinfo{author}{\bibfnamefont{M.}~\bibnamefont{Inoue}},
  \bibinfo{journal}{Prog. Theor. Phys.} \textbf{\bibinfo{volume}{78}},
  \bibinfo{pages}{787} (\bibinfo{year}{1987}).

\bibitem[{\citenamefont{Kl\"umper}(1992)}]{kl92rsos}
\bibinfo{author}{\bibfnamefont{A.}~\bibnamefont{Kl\"umper}},
  \bibinfo{journal}{Ann. Phys. Lpz.} \textbf{\bibinfo{volume}{1}},
  \bibinfo{pages}{540} (\bibinfo{year}{1992}).

\bibitem[{\citenamefont{Kl\"umper}(1993)}]{kl93}
\bibinfo{author}{\bibfnamefont{A.}~\bibnamefont{Kl\"umper}},
  \bibinfo{journal}{Z. Phys. B} \textbf{\bibinfo{volume}{91}},
  \bibinfo{pages}{507} (\bibinfo{year}{1993}).

\bibitem[{\citenamefont{Korepin}(1982)}]{kor82}
\bibinfo{author}{\bibfnamefont{V.~E.} \bibnamefont{Korepin}},
  \bibinfo{journal}{Comm. Math. Phys.} \textbf{\bibinfo{volume}{86}},
  \bibinfo{pages}{391} (\bibinfo{year}{1982}).

\bibitem[{\citenamefont{Slavnov}(1989)}]{sla89}
\bibinfo{author}{\bibfnamefont{N.~A.} \bibnamefont{Slavnov}},
  \bibinfo{journal}{Teor. Mat. Fiz.} \textbf{\bibinfo{volume}{79}},
  \bibinfo{pages}{232} (\bibinfo{year}{1989}).

\bibitem[{\citenamefont{Kitanine et~al.}(2000)\citenamefont{Kitanine, Maillet,
  and Terras}}]{kit00}
\bibinfo{author}{\bibfnamefont{N.}~\bibnamefont{Kitanine}},
  \bibinfo{author}{\bibfnamefont{M.~J.} \bibnamefont{Maillet}},
  \bibnamefont{and} \bibinfo{author}{\bibfnamefont{V.}~\bibnamefont{Terras}},
  \bibinfo{journal}{Nucl. Phys. B} \textbf{\bibinfo{volume}{567}},
  \bibinfo{pages}{554} (\bibinfo{year}{2000}).

\bibitem[{\citenamefont{Kitanine et~al.}(2002)\citenamefont{Kitanine, Maillet,
  Slavnov, and Terras}}]{kit02}
\bibinfo{author}{\bibfnamefont{N.}~\bibnamefont{Kitanine}},
  \bibinfo{author}{\bibfnamefont{M.~J.} \bibnamefont{Maillet}},
  \bibinfo{author}{\bibfnamefont{N.~A.} \bibnamefont{Slavnov}},
  \bibnamefont{and} \bibinfo{author}{\bibfnamefont{V.}~\bibnamefont{Terras}},
  \bibinfo{journal}{Nucl. Phys. B} \textbf{\bibinfo{volume}{641}},
  \bibinfo{pages}{487} (\bibinfo{year}{2002}).

\bibitem[{\citenamefont{Tsuboi and Shiroishi}(2005)}]{tsu05}
\bibinfo{author}{\bibfnamefont{Z.}~\bibnamefont{Tsuboi}} \bibnamefont{and}
  \bibinfo{author}{\bibfnamefont{M.}~\bibnamefont{Shiroishi}},
  \bibinfo{journal}{J. Phys. A} \textbf{\bibinfo{volume}{38}}, \bibinfo{pages}{L363} (\bibinfo{year}{2005}).

\bibitem[{\citenamefont{Takahashi}(1977)}]{tak77}
\bibinfo{author}{\bibfnamefont{M.}~\bibnamefont{Takahashi}},
  \bibinfo{journal}{J. Phys. C} \textbf{\bibinfo{volume}{10}},
  \bibinfo{pages}{1289} (\bibinfo{year}{1977}).

\bibitem[{\citenamefont{Boos and Korepin}(2001)}]{boo01}
\bibinfo{author}{\bibfnamefont{H.}~\bibnamefont{Boos}} \bibnamefont{and}
  \bibinfo{author}{\bibfnamefont{V.~E.} \bibnamefont{Korepin}},
  \bibinfo{journal}{J. Phys. A} \textbf{\bibinfo{volume}{34}},
  \bibinfo{pages}{5311} (\bibinfo{year}{2001}).

\bibitem[{\citenamefont{Kato et~al.}(2003)\citenamefont{Kato, Shiroishi,
  Takahashi, and Sakai}}]{kat03}
\bibinfo{author}{\bibfnamefont{G.}~\bibnamefont{Kato}},
  \bibinfo{author}{\bibfnamefont{M.}~\bibnamefont{Shiroishi}},
  \bibinfo{author}{\bibfnamefont{M.}~\bibnamefont{Takahashi}},
  \bibnamefont{and} \bibinfo{author}{\bibfnamefont{K.}~\bibnamefont{Sakai}},
  \bibinfo{journal}{J. Phys. A} \textbf{\bibinfo{volume}{36}},
  \bibinfo{pages}{L337} (\bibinfo{year}{2003}).

\bibitem[{\citenamefont{Kato et~al.}(2004)\citenamefont{Kato, Shiroishi,
  Takahashi, and Sakai}}]{kat04}
\bibinfo{author}{\bibfnamefont{G.}~\bibnamefont{Kato}},
  \bibinfo{author}{\bibfnamefont{M.}~\bibnamefont{Shiroishi}},
  \bibinfo{author}{\bibfnamefont{M.}~\bibnamefont{Takahashi}},
  \bibnamefont{and} \bibinfo{author}{\bibfnamefont{K.}~\bibnamefont{Sakai}},
  \bibinfo{journal}{J. Phys. A} \textbf{\bibinfo{volume}{37}},
  \bibinfo{pages}{5097} (\bibinfo{year}{2004}).

\bibitem[{\citenamefont{Boos et~al.}(2003)\citenamefont{Boos, Korepin, and
  Smirnov}}]{boo03}
\bibinfo{author}{\bibfnamefont{H.}~\bibnamefont{Boos}},
  \bibinfo{author}{\bibfnamefont{V.}~\bibnamefont{Korepin}}, \bibnamefont{and}
  \bibinfo{author}{\bibfnamefont{F.}~\bibnamefont{Smirnov}},
  \bibinfo{journal}{Nucl. Phys. B} \textbf{\bibinfo{volume}{658}},
  \bibinfo{pages}{417} (\bibinfo{year}{2003}).

\bibitem[{\citenamefont{Boos et~al.}(2005)\citenamefont{Boos, Shiroishi, and
  Takahashi}}]{boo05}
\bibinfo{author}{\bibfnamefont{H.}~\bibnamefont{Boos}},
  \bibinfo{author}{\bibfnamefont{M.}~\bibnamefont{Shiroishi}},
  \bibnamefont{and}
  \bibinfo{author}{\bibfnamefont{M.}~\bibnamefont{Takahashi}},
  \bibinfo{journal}{Nucl. Phys. B} \textbf{\bibinfo{volume}{712}},
  \bibinfo{pages}{573} (\bibinfo{year}{2005}).

\bibitem[{\citenamefont{McCoy}(1968)}]{mcc68}
\bibinfo{author}{\bibfnamefont{B.~M.} \bibnamefont{McCoy}},
  \bibinfo{journal}{Phys. Rev.} \textbf{\bibinfo{volume}{173}},
  \bibinfo{pages}{531} (\bibinfo{year}{1968}).

\bibitem[{\citenamefont{Shiroishi et~al.}(2001)\citenamefont{Shiroishi,
  Takahashi, and Nishiyama}}]{shi01}
\bibinfo{author}{\bibfnamefont{M.}~\bibnamefont{Shiroishi}},
  \bibinfo{author}{\bibfnamefont{M.}~\bibnamefont{Takahashi}},
  \bibnamefont{and}
  \bibinfo{author}{\bibfnamefont{Y.}~\bibnamefont{Nishiyama}},
  \bibinfo{journal}{J. Phys. Soc. Jap.} \textbf{\bibinfo{volume}{70}},
  \bibinfo{pages}{3535} (\bibinfo{year}{2001}).

\bibitem[{\citenamefont{Baxter}(1982)}]{bax82}
\bibinfo{author}{\bibfnamefont{R.}~\bibnamefont{Baxter}},
  \emph{\bibinfo{title}{Exactly {Solved Models in Statistical Mechanics}}}
  (\bibinfo{publisher}{Academic Press London}, \bibinfo{year}{1982}).

\bibitem[{\citenamefont{Essler et~al.}(2005)\citenamefont{Essler, Frahm,
  G\"ohmann, Kl\"umper, and Korepin}}]{book}
\bibinfo{author}{\bibfnamefont{F.~H.~L.} \bibnamefont{Essler}},
  \bibinfo{author}{\bibfnamefont{H.}~\bibnamefont{Frahm}},
  \bibinfo{author}{\bibfnamefont{F.}~\bibnamefont{G\"ohmann}},
  \bibinfo{author}{\bibfnamefont{A.}~\bibnamefont{Kl\"umper}},
  \bibnamefont{and} \bibinfo{author}{\bibfnamefont{V.~E.}
  \bibnamefont{Korepin}}, \emph{\bibinfo{title}{The {O}ne-dimensional {H}ubbard
  Model}} (\bibinfo{publisher}{Cambridge University Press},
  \bibinfo{year}{2005}).

\bibitem[{\citenamefont{Jimbo et~al.}(1992)\citenamefont{Jimbo, Miki, Miwa, and
  Nakayashiki}}]{jim92}
\bibinfo{author}{\bibfnamefont{M.}~\bibnamefont{Jimbo}},
  \bibinfo{author}{\bibfnamefont{K.}~\bibnamefont{Miki}},
  \bibinfo{author}{\bibfnamefont{T.}~\bibnamefont{Miwa}}, \bibnamefont{and}
  \bibinfo{author}{\bibfnamefont{A.}~\bibnamefont{Nakayashiki}},
  \bibinfo{journal}{Phys. Lett. A} \textbf{\bibinfo{volume}{168}},
  \bibinfo{pages}{256} (\bibinfo{year}{1992}).

\bibitem[{\citenamefont{Jimbo and Miwa}(1996)}]{jm96}
\bibinfo{author}{\bibfnamefont{M.}~\bibnamefont{Jimbo}} \bibnamefont{and}
  \bibinfo{author}{\bibfnamefont{T.}~\bibnamefont{Miwa}}, \bibinfo{journal}{J.
  Phys. A} \textbf{\bibinfo{volume}{29}}, \bibinfo{pages}{2923}
  (\bibinfo{year}{1996}).

\bibitem[{\citenamefont{Whittacker and Watson}(1935)}]{ww}
\bibinfo{author}{\bibfnamefont{E.~T.} \bibnamefont{Whittacker}}
  \bibnamefont{and} \bibinfo{author}{\bibfnamefont{G.~N.}
  \bibnamefont{Watson}}, \emph{\bibinfo{title}{{A Course of Modern Analysis}}}
  (\bibinfo{publisher}{Cambridge University Press}, \bibinfo{year}{1935}).

\bibitem[{\citenamefont{Kl\"umper}(1998)}]{klpei}
\bibinfo{author}{\bibfnamefont{A.}~\bibnamefont{Kl\"umper}},
  \bibinfo{journal}{Eur. Phys. J. B} \textbf{\bibinfo{volume}{5}},
  \bibinfo{pages}{677} (\bibinfo{year}{1998}).

\bibitem[{\citenamefont{Johnson et~al.}(1973)\citenamefont{Johnson, Krinsky,
  and McCoy}}]{joh73}
\bibinfo{author}{\bibfnamefont{J.~D.} \bibnamefont{Johnson}},
  \bibinfo{author}{\bibfnamefont{S.}~\bibnamefont{Krinsky}}, \bibnamefont{and}
  \bibinfo{author}{\bibfnamefont{B.~M.} \bibnamefont{McCoy}},
  \bibinfo{journal}{Phys. Rev. A} \textbf{\bibinfo{volume}{8}},
  \bibinfo{pages}{2526} (\bibinfo{year}{1973}).

\bibitem[{\citenamefont{Fabricius et~al.}(1998)\citenamefont{Fabricius,
  Kl\"umper, and McCoy}}]{fab98}
\bibinfo{author}{\bibfnamefont{K.}~\bibnamefont{Fabricius}},
  \bibinfo{author}{\bibfnamefont{A.}~\bibnamefont{Kl\"umper}},
  \bibnamefont{and} \bibinfo{author}{\bibfnamefont{B.~M.} \bibnamefont{McCoy}},
  \bibinfo{journal}{arXiv:cond-mat/9810278}  (\bibinfo{year}{1998}).

\bibitem[{\citenamefont{Johnson and McCoy}(1972)}]{joh72}
\bibinfo{author}{\bibfnamefont{J.~D.} \bibnamefont{Johnson}} \bibnamefont{and}
  \bibinfo{author}{\bibfnamefont{B.~M.} \bibnamefont{McCoy}},
  \bibinfo{journal}{Phys. Rev. A} \textbf{\bibinfo{volume}{6}},
  \bibinfo{pages}{1613} (\bibinfo{year}{1972}).

\bibitem[{\citenamefont{Suzuki}(2002)}]{suz02}
\bibinfo{author}{\bibfnamefont{M.} \bibnamefont{Suzuki}},
  \bibinfo{journal}{Phys. Lett. A} \textbf{\bibinfo{volume}{301}},
  \bibinfo{pages}{398} (\bibinfo{year}{2002}).

\bibitem[{\citenamefont{G\"ohmann and Seel}(2005)}]{gs05}
\bibinfo{author}{\bibfnamefont{F.}~\bibnamefont{G\"ohmann}} \bibnamefont{and}
  \bibinfo{author}{\bibfnamefont{A.}~\bibnamefont{Seel}}, \bibinfo{journal}{arXiv:hep-th/0505091}  (\bibinfo{year}{2005}).

\bibitem[{\citenamefont{Boos et~al.}(2004{\natexlab{a}})\citenamefont{Boos,
  Jimbo, Miwa, Smirnov, and Takeyama}}]{boo04}
\bibinfo{author}{\bibfnamefont{H.}~\bibnamefont{Boos}},
  \bibinfo{author}{\bibfnamefont{M.}~\bibnamefont{Jimbo}},
  \bibinfo{author}{\bibfnamefont{T.}~\bibnamefont{Miwa}},
  \bibinfo{author}{\bibfnamefont{F.}~\bibnamefont{Smirnov}}, \bibnamefont{and}
  \bibinfo{author}{\bibfnamefont{Y.}~\bibnamefont{Takeyama}},
  \bibinfo{journal}{hep-th/0405044}  (\bibinfo{year}{2004}{\natexlab{a}}).

\bibitem[{\citenamefont{Boos et~al.}(2004{\natexlab{b}})\citenamefont{Boos,
  Jimbo, Miwa, Smirnov, and Takeyama}}]{boo04b}
\bibinfo{author}{\bibfnamefont{H.}~\bibnamefont{Boos}},
  \bibinfo{author}{\bibfnamefont{M.}~\bibnamefont{Jimbo}},
  \bibinfo{author}{\bibfnamefont{T.}~\bibnamefont{Miwa}},
  \bibinfo{author}{\bibfnamefont{F.}~\bibnamefont{Smirnov}}, \bibnamefont{and}
  \bibinfo{author}{\bibfnamefont{Y.}~\bibnamefont{Takeyama}},
  \bibinfo{journal}{hep-th/0412191}  (\bibinfo{year}{2004}{\natexlab{b}}).

\bibitem[{\citenamefont{G\"ohmann et~al.}(2005)\citenamefont{G\"ohmann,
  Hasenclever, and Seel}}]{has05}
\bibinfo{author}{\bibfnamefont{F.}~\bibnamefont{G\"ohmann}},
  \bibinfo{author}{\bibfnamefont{N.~P.}~\bibnamefont{Hasenclever}},
  \bibnamefont{and} \bibinfo{author}{\bibfnamefont{A.}~\bibnamefont{Seel}},
  \bibinfo{journal}{in preparation} (\bibinfo{year}{2005}).

\end{thebibliography}
\end{document}